\documentclass[aps,pra,twocolumn,showpacs]{revtex4}
\usepackage{graphicx}
\usepackage{amssymb}
\usepackage{amsmath}
\usepackage{bm}

\begin{document}

\title{Relaxation of superfluid turbulence in highly oblate Bose-Einstein condensates}
\author{Woo Jin Kwon}
\author{Geol Moon}
\author{Jae-yoon Choi}
\author{Sang Won Seo}
\author{Yong-il Shin}
\email{yishin@snu.ac.kr}

\affiliation{Center for Subwavelength Optics and Department of Physics and Astronomy, Seoul National University, Seoul 151-747, Korea}

\begin{abstract}
We investigate thermal relaxation of superfluid turbulence in a highly oblate Bose-Einstein condensate. We generate turbulent flow in the condensate by sweeping the center region of the condensate with a repulsive optical potential. The turbulent condensate shows a spatially disordered distribution of quantized vortices and  the vortex number of the condensate exhibits nonexponential decay behavior which we attribute to the vortex pair annihilation. The vortex-antivortex collisions in the condensate are identified with crescent-shaped, coalesced vortex cores. We observe that the nonexponential decay of the vortex number is quantitatively well described by a rate equation consisting of one-body and two-body decay terms. In our measurement, we find that the local two-body decay rate is closely proportional to $T^2/\mu$, where $T$ is the temperature and $\mu$ is the chemical potential.
\end{abstract}


\maketitle

\section{Introduction}
In a two-dimensional (2D) superfluid, quantized vortices are topological, point-like objects and they can be created and also annihilated as a pair of vortices of opposite circulation. Vortex-antivortex pairs play essential roles in 2D superfluid phenomena such as the Berezinskii-Kosterlitz-Thouless transition~\cite{Berezinskii,KT}, phase transition dynamics~\cite{Zurek}, and superfluid turbulence~\cite{onorato,Tsubota_review}. Recently, controlled experimental studies of vortex dipole dynamics have been enabled in atomic Bose-Einstein condensate (BEC) systems~\cite{neely,hall} and thermal activation of vortex pairs has been observed in quasi-2D Bose gases~\cite{Hadzibabic,choi1}. However, the annihilation of a vortex-antivortex pair has not been clearly observed yet.

Vortex pair annihilation is of particular importance in 2D superfluid turbulence. In 2D turbulence of a classical hydrodynamic fluid, the kinetic energy of the system flows toward large length scales due to the conservation of enstrophy which is integral of squared vorticity~\cite{kraichnan_review}. This is known as the inverse energy cascade and manifests itself in generating large-scale flow structures from small-scale forcing. This phenomenon is qualitatively different from three-dimensional (3D) turbulence where energy is typically dissipated into small length scales. An interesting question is whether the inverse cascade can occur in an atomic BEC. Since the enstrophy, proportional to the total number of quantized vortices in quantum turbulence, is not conserved in a compressible 2D superfluid due to the vortex-antivortex annihilation, there has been a theoretical controversy on this issue~\cite{horng,tsubota1,tsubota2,nowak1,nowak2,takuya,nowak,reeves1,reeves2,holographic}. Recently, Neely \textit{et al.}~\cite{anderson} reported an experimental and numerical study to show that there are conditions for which 2D turbulence in a BEC can dissipatively evolve into large-scale flow.

In this paper, we investigate thermal relaxation of turbulent superflow in highly oblate BECs.  By sweeping the center region of a trapped condensate with a repulsive optical potential, we generate turbulent flow with a spatially disordered vortex distribution.  We measure the temporal evolution of the vortex number and observe nonexponential decay behavior in the relaxation, which we attribute to the vortex pair annihilation. The vortex-antivortex collisions in the condensate are identified with crescent-shaped, coalesced vortex cores. We characterize the nonexponential decay of the vortex number with one-body and two-body decay rates, and find in our measurements that the local two-body decay rate is closely proportional to $T^2/\mu$, where $T$ is the temperature and $\mu$ is the chemical potential of the sample. Our results on the decay rates provide quantitative information on the thermal dissipation in 2D quantum turbulence, in particualr, with finite compressibility.

This paper is organized as follows. In Sec.~II, we describe our experimental setup and the experimental procedure for generating turbulent flow in the condensate. In Sec.~III, we present the experimental observations on the vortex-antivortex collision and the analysis of the nonexponential decay of the vortex number of the turbulent condensate. Finally, a summary is presented in Sev.~IV.

\section{Experiment}

\begin{figure}
\includegraphics[width=8cm]{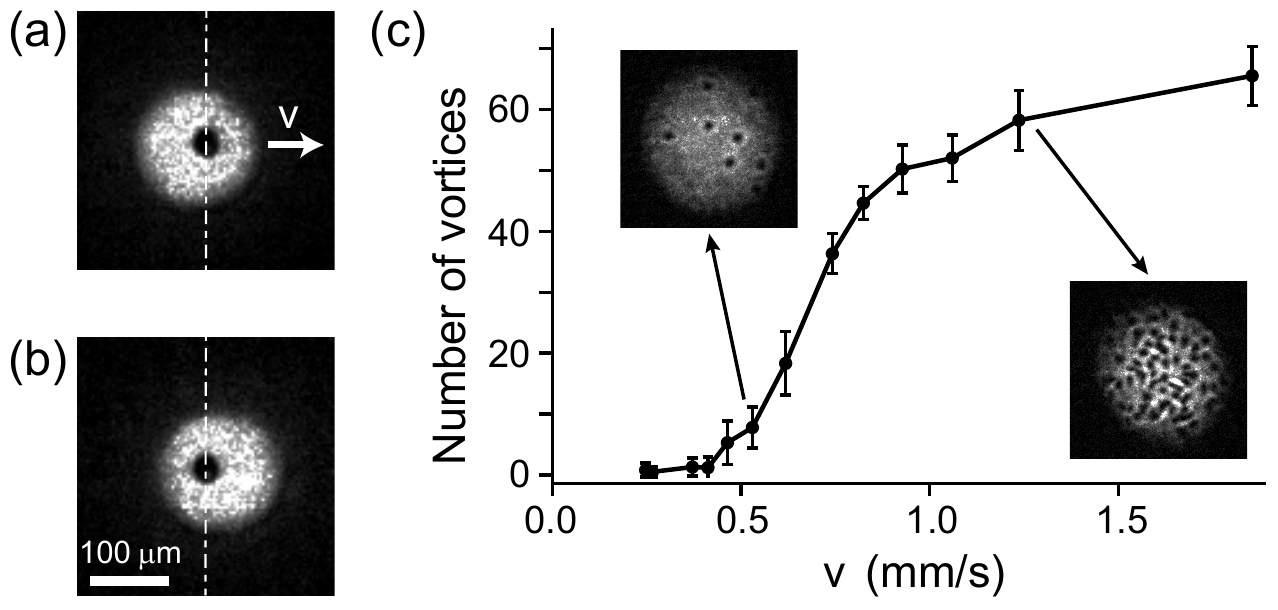}
\caption{Generation of turbulent flow in a BEC. A repulsive laser beam penetrates through the condensate and sweeps its center region by horizontally translating the trapped condensate by 37~$\mu$m. Images of the trapped condensate (a) before and (b) after the translation. (c) Number of the vortices in the perturbed condensate as a function of the translation speed $v$. The speed of sound is estimated to be $c\approx 4.6$~mm/s at the center of the condensate.}
\label{preparation}
\end{figure}

We prepare a highly oblate BEC of $^{23}\textrm{Na}$ atoms in the $|F=1,m_{F}=-1\rangle$ state in a harmonic trap, where the axial and radial confinements are provided by optical and magnetic trapping potentials, respectively. The magnetic potential is generated from an axially symmetric, magnetic quadrupole field~\cite{choi1}. The trapping frequencies are $\omega_{r,z}=2\pi \times(15,350)$~Hz. For a condensate of $N_0=1.8\times 10^6$ atoms, the chemical potential is $\mu\approx k_B \times 60$~nK and the radial Thomas-Fermi radius is $R=\sqrt{2\mu/m\omega_r^2}\approx 70~\mu$m, where $m$ is the atomic mass. Because of the large aspect ratio of the condensate, $\omega_z/\omega_r>20$, the vortex line excitations are highly suppressed~\cite{Jackson09,rooney} and we expect that the vortex dynamics in our system is 2D. Note that $\mu>3\hbar\omega_z$ and the condensate is thermodynamically 3D.

To generate turbulent flow in the condensate, we employ a repulsive Gaussian laser beam as a stirring obstacle~\cite{neely}. Having the laser beam axially penetrating through the condensate, we translate the condensate in a transverse direction by moving the magnetic trap center by 37~$\mu$m for 30~ms (Fig.~1), and adiabatically turn off the laser beam for 0.4~s. Here the laser beam, fixed in the lab frame, sweeps the center region of the condensate. The Gaussian beam waist is $15~\mu$m and the barrier height is about 15$\mu$. The translation speed corresponds to $\sim0.3 c$, where $c=\sqrt{\mu/m}$ is the speed of sound in the center region of the condensate. With this procedure, we could generate over 60 vortices in our coldest sample without inducing noticeable shape oscillations of the condensate. The critical velocity for the vortex nucleation was measured to be $\approx 0.1c$ [Fig.~1(c)].

We detect vortices in the turbulent flow by taking an absorption image after 24~ms time-of-flight. In releasing the tapping potential, we turn off the magnetic potential 12~ms earlier than the optical potential. This was found to be helpful to improve the vortex core visibility because the axial confinement direction of the optical trap was well aligned to the imaging axis~\cite{choi1}.

\begin{figure}
\includegraphics[width=8.5cm]{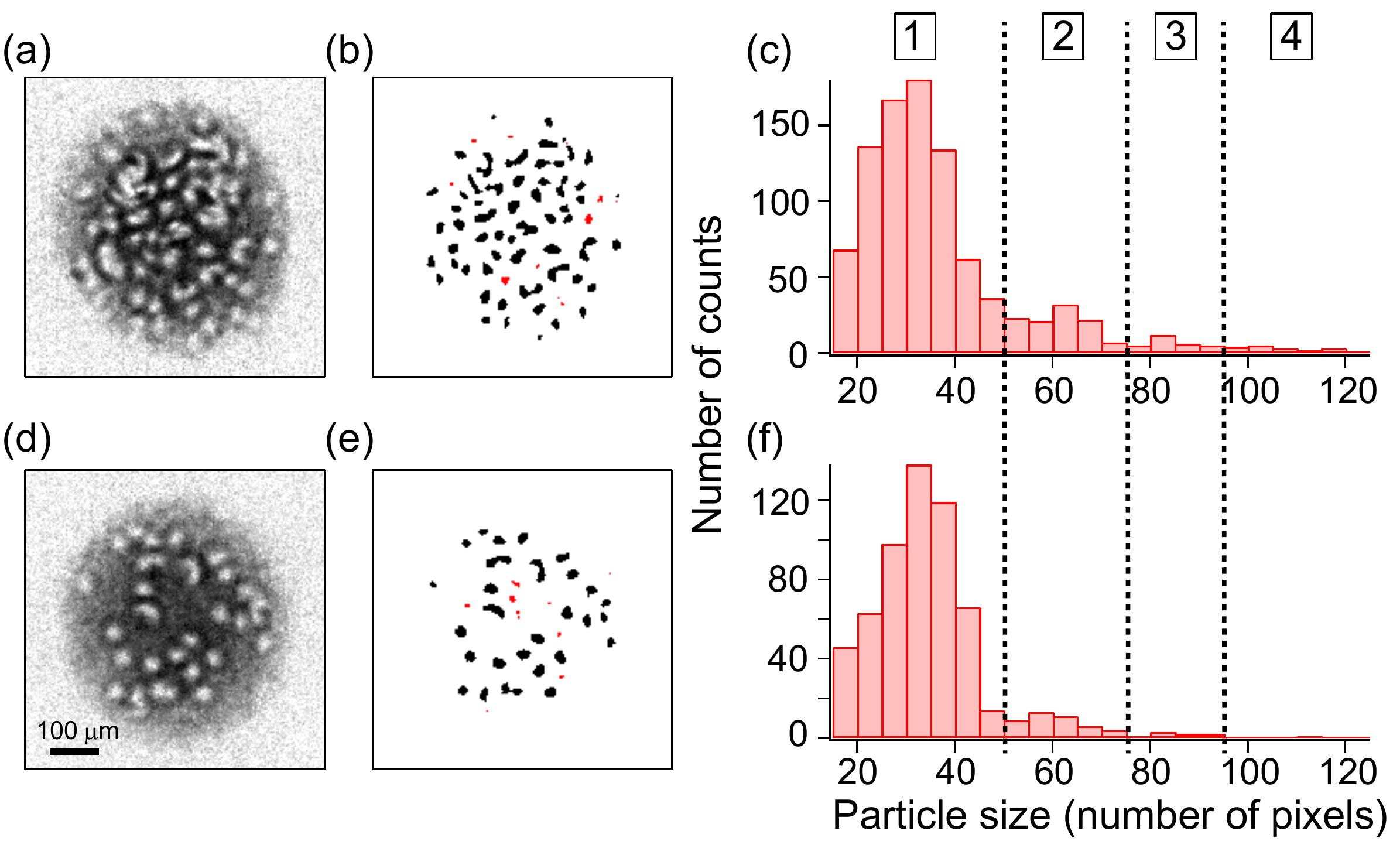}
\caption{(Color online) Vortex number counting for a turbulent condensate with a large number of vortices. (a) Absorption image of the condensate after expansion. (b) A binary image obtained from (a) and its blurred one (see the text). Black particles correspond to the vortex cores and  red particles, having area sizes less than 15 pixels, are the artifacts due to image defects. (c) Histogram of the particle area size from 24 image data. We set the vortex number transition lines (dashed lines) in the middle of the peaks. (d)-(f) display another example set of the image analysis, where the condensate contains about 30 vortices.}
\label{Fig2}
\end{figure}

An image processing method is developed to facilitate measuring the number of vortices, $N_v$, in a turbulent sample (Fig.~\ref{Fig2}). First, we produce a blurred image by applying boxcar smoothing to an absorption image, where the box width is set to be 30~$\mu$m, comparable to the vortex core diameter in the image. We divide the original absorption image by its blurred image and then convert it into a binary image for a certain threshold value. With this image processing, the density-depleted vortex cores in the absorption image are transformed into particles in the binary image [Fig.~\ref{Fig2}(b) and (e)]. Each particle has a different area size depending on the number of vortices it has. The histogram of the particle area size shows a multiple peak structure. We set the vortex number transition lines in the middle of the peaks and assign a vortex number to each particle [Fig.~\ref{Fig2}(c) and (f)]. The particles having an area size less than 15 pixels are due to image defects and ignored in our counting.

In order to validate this $N_v$ counting method, we compared its results with the results obtained from hand counting for many images and confirmed that the image processing method gives a consistent counting within less than $10\%$ ($\sim5$ vortices). The positions of the vortex number transition lines do not significantly affect the counting because the relative number of the particles having many vortices is small. Most of uncertainty in the counting was found to come from the vortices in the the boundary region. In our analysis, we used the image processing method for the images with a large number of vortices and for $N_v<30$, we counted vortices by hand.

\section{Results}

\subsection{Nonexponential decay}

\begin{figure*}
\includegraphics[width=14.5cm]{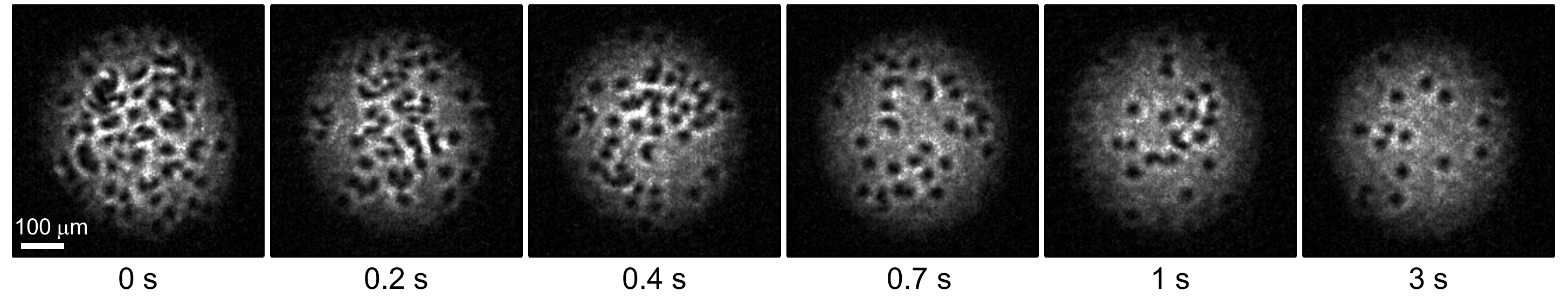}
\caption{Relaxation of turbulent supeflow in a highly oblate BEC. Examples of images for various relaxation times, where the atom number of the condensate $N_0\approx1.8\times10^{6}$ and the condensate fraction $\eta\approx 0.8$.}
\label{decay image}
\end{figure*}

Figure~\ref{decay image} displays images of condensates with turbulent flow after various relaxation times. The turbulent condensate shows spatially disordered distributions of vortices.  In the early phase of the evolution, some features in the condensate seem to be suggestive of vortex clustering, but we observed no evidence in the spatial analysis of the vortex distributions using Ripley's \textit{K}-function~\cite{white}. The turbulent condensate eventually relaxes into a stationary state as the vortex number decreases.

\begin{figure}
\includegraphics[width=7.0cm]{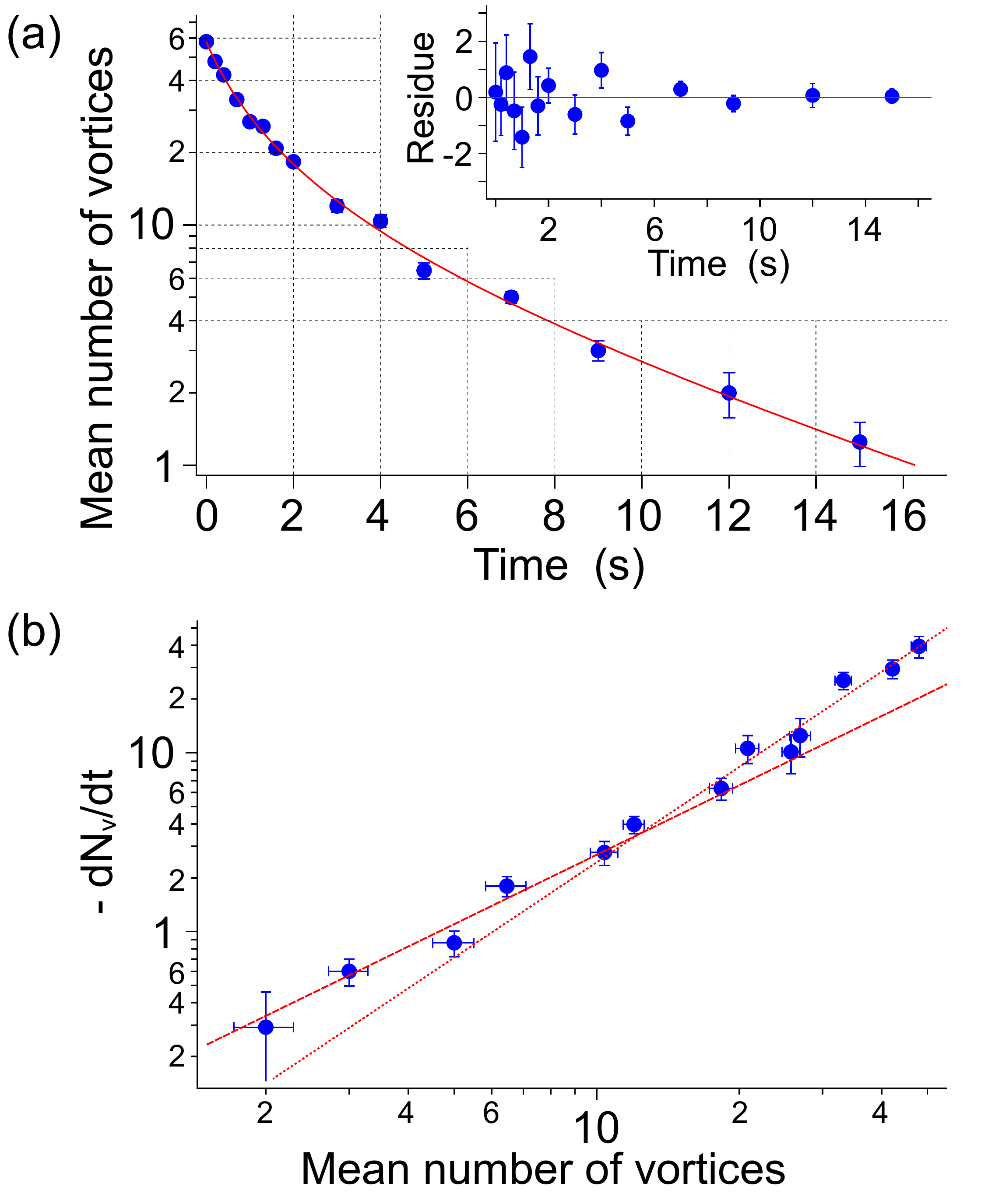}
\caption{(Color online) Nonexponential decay of the vortex number. (a) Mean number of vortices versus relaxation time. The sample condition is the same as in Fig.~3. Each measurement point consists of at least 12 realizations of the same experiment and the error bars indicate the standard deviations of the mean vortex number. The solid line denotes a fit of the rate equation in Eq.~(1) to the data. The inset shows the residue of the data from the fitting line. (b) Decay rate of the mean vortex number, $-dN_v /dt$, as a function of $N_v$. The dashed line and the dotted line are power-law fitting lines to the data points for $N_v<10$ and $N_v>10$, giving the exponents of 1.29(17) and 1.77(20), respectively. }
\label{nonexpon}
\end{figure}

We observe nonexponential decay behavior of the vortex number in the relaxation~[Fig.~\ref{nonexpon}(a)], where the decay rate is faster for higher vortex number. There are only two ways for a quantum vortex to disappear from a condensate of finite spatial extent: drifting out of the condensate or being annihilated as a vortex-antivortex pair inside the condensate. Since our turbulence generation method imparts no angular momentum to the condensate, the initial turbulent condensate would have equal numbers of vortices for clockwise and counterclockwise circulations, and both of the drifting-out and the pair annihilation processes must be involved in the relaxation dynamics. Since the pair annihilation is intrinsically a two-vortex process, it can be a reason for the nonexponential decay of the vortex number.

To examine the details of the decay behavior of $N_v$, we determine the decay rate $-dN_v /dt$ from the measured $N_v$ data and Fig.~\ref{nonexpon}(b) displays it as a function of $N_v$. In the log-log plot, it is clearly seen that the dependence of the decay rate on $N_v$ cannot be captured by a single power-law relation over the whole range of our measurements. The power-law fits for the low vortex number ($N_v<10$) data and the high vortex number ($N_v>10$) data give the exponents of 1.29(17) and 1.77(20), respectively. 

Partly motivated by the measured values of the exponents, we suggest a phenomenological rate equation for $N_v$ as
\begin{equation}
\frac{dN_v }{dt }= - \Gamma_1 N_v -\Gamma_2 N_v^2, 
\end{equation}
and observe that the measured decay curve of $N_v$ is remarkably well described with this rate equation. The residue of the experiment data from the fitting line of the rate equation is smaller than 2 vortices~[Fig.~\ref{nonexpon}(a) inset]. A simple consideration based on the kinetic gas theory suggests that the one-body and two-body decay rates, $\Gamma_1$ and $\Gamma_2$, might be mainly determined by the drifting-out process and the pair annihilation process, respectively. However, we cannot exclude many-vortex effects on the drifting-out process, which would possibly affect the two-body decay rate $\Gamma_2$. To the best of our knowledge, there is no predicted form for the decay curve of $N_v$ in a turbulent trapped condensate, taking both of the drifting-out and pair annihilation processes into account. In this work, we employ the two decay rates, $\Gamma_1$ and $\Gamma_2$, to characterize the relaxation of the turbulent condensate. 

\subsection{Vortex-antivortex collision}

\begin{figure}
\includegraphics[width=7.0cm]{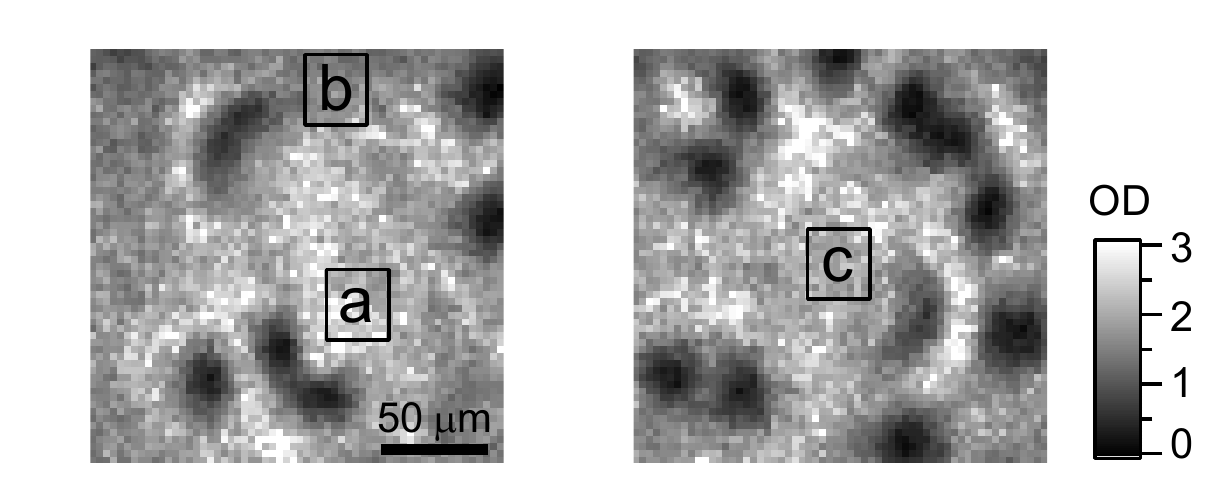}
\caption{Vortex pair annihilation in the turbulent superflow. Density-depleted regions with a crescent shape are observed in the turbulent condensates. In a vortex-antivortex collision event, two vortex cores can coalesce (a), evolve into a dark soliton (b), and disappear as atoms fill up the density-depleted region (c). The bending structure results from the linear momentum of the vortex dipole.}
\label{annihilation image}
\end{figure}

Vortex pair annihilation occurs when two vortices of opposite circulation collide, converting their energy into sound waves in the superfluid. Numerical studies showed that a dark or gray soliton of a crescent shape is formed via coalescing the vortex cores in the collision and it can dissipatively evolve into a shock wave~\cite{onorato,rorari13,angom}. Indeed, we observe crescent-shaped density-depleted regions in the condensate (Fig.~\ref{annihilation image}), revealing the vortex-antivortex collision events in the turbulent flow. The bending structure can be accounted for by the linear momentum of the vortex dipole which is perpendicular to the vortex dipole direction. Because of the movement, the atomic density on the convex side of the vortex dipole is higher than that in the opposite side. Some of the coalesced vortex cores appear with significantly reduced visibility [Fig.~\ref{annihilation image}(c)], possibly indicating that they are being annihilated.

We emphasize that such a crescent-shaped vortex core was not observed in rotating, turbulent condensates~\cite{choi,seji}. In this case, turbulent flow was generated by circularly shaking the magnetic trapping potential, where most of the vortices have same circulation. The trap geometry and the imaging procedure were identical to those in this work, thus excluding the possibility of imaging artifacts such as vortex line tilting. The crescent-shaped density dimple obviously comes from a vortex-antivortex pair.

\begin{figure}
\includegraphics[width=7.3cm]{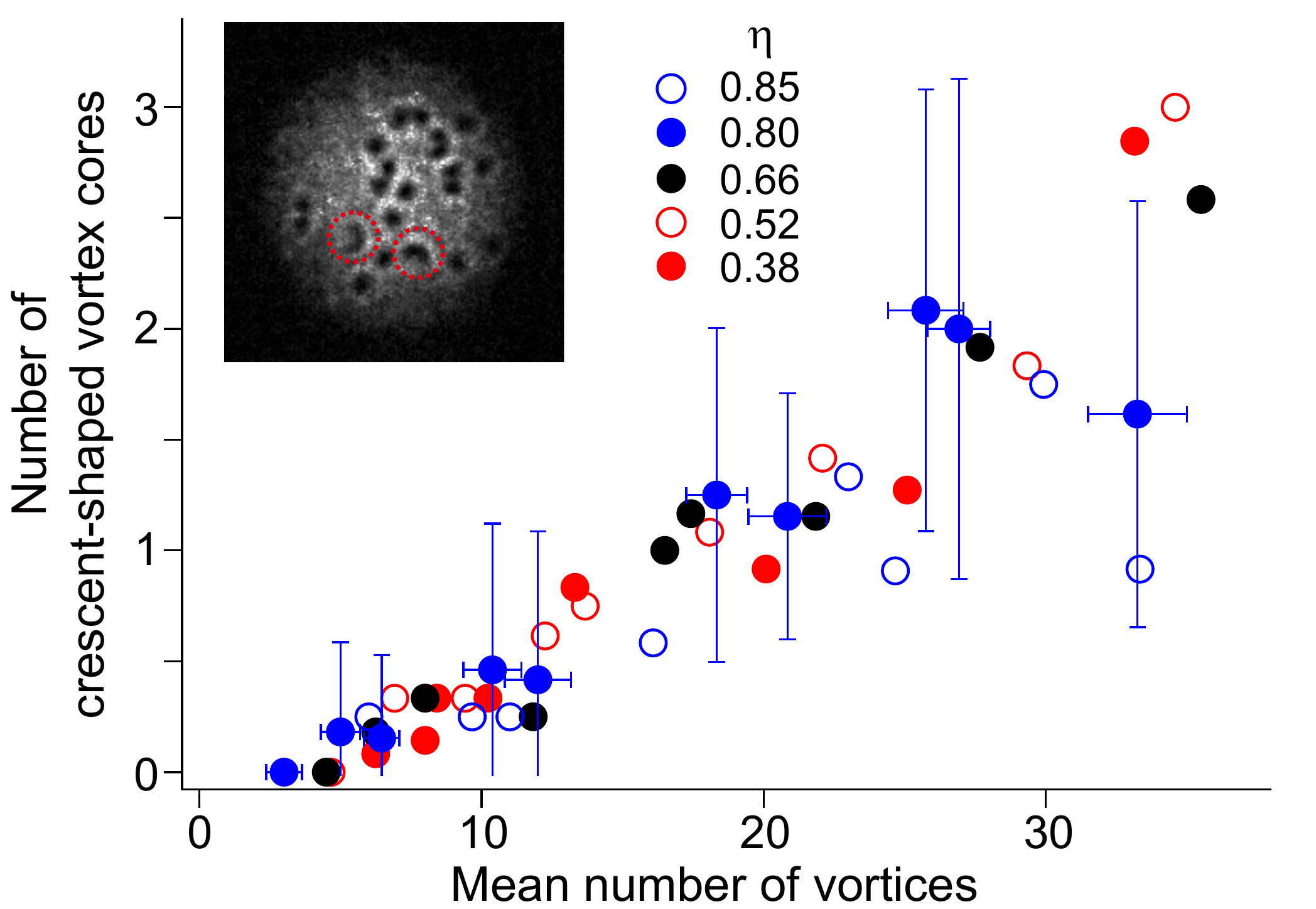}
\caption{(Color online) The number of crescent-shaped vortex cores versus the total vortex number $N_v$ in a turbulent condensate. Each measurement point consists of at least 12 realizations of the same experiment. The inset shows an image of a turbulent condensate with two cresent-shaped vortex cores (red dashed circles).}
\label{freq}
\end{figure}

The appearance frequency of the crescent-shaped density dimple is found to be almost linearly proportional to the total vortex number $N_v$ and insensitive to the sample temperature~(Fig.~\ref{freq}). At high vortex density $n_v$, a 2D turbulent superfluid can be considered as a gas of vortex-antivortex pairs~\cite{takuya,nowak}. For a vortex pair of size $d$, the collisional cross section $\sigma\sim d$ and the linear velocity $v\sim \hbar/m d $, giving the vortex collision rate as $\gamma_c=\sigma v n_v/2\sim (\hbar/2m) n_v$. This estimation provides a qualitative explanation of the observed $N_v$ dependence of the appearance frequency of the crescent-shaped density dimple. We note that the numerical results in Ref.~\cite{takuya,nowak} for a homogeneous system showed that $\gamma_c$ has different power-law depedence on $n_v$ at low vortex density, $n_v \xi^2 <3\times10^{-3}$ ($\xi=\hbar/\sqrt{2m \mu}$ is the vortex core size). The vortex density of our sample is $n_v\xi^2< 10^{-3}$ with $n_v=N_v/(\pi R^2)$, but direct comparison of our observation to the prediction is limited due to the sample inhomogeneity.

It should be pointed out that a vortex-antivortex collision event does not necessarily result in pair annihilation. A tightly bound pair, having a high linear momentum, would cross the condensate to the boundary region and the pair might have a chance to move out of the condensate with the aid of a certain energy dissipation mechanism.

\subsection{Decay rate measurements}

 The local vortex dynamics in a homogeneous system is governed by the temperature $T$ and the chemical potential $\mu$ of the system: at finite temperature, a vortex experiences a friction force caused by collisional exchange of atoms between the condensate and the thermal cloud~\cite{HV,thermalQT,vring}, and the chemical potential determines the vortex core size $\xi\propto\mu^{-1/2}$, providing a characteristic length scale in the vortex dynamics. In this subsection, we present the measurement results of the decay rates of the vortex number for various sample conditions.

\subsubsection{Temperature}

\begin{figure}
\includegraphics[width=8.5cm]{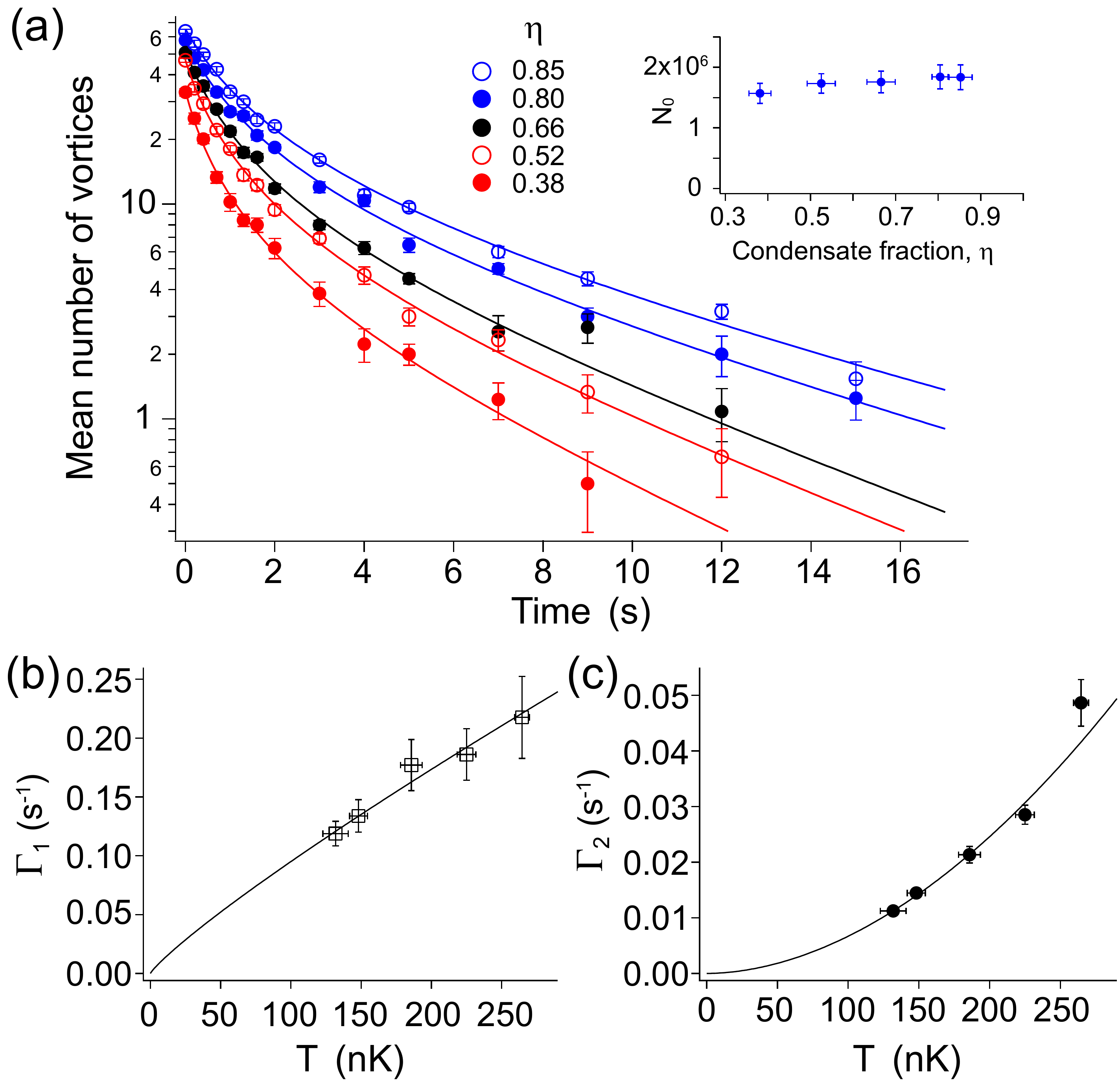}
\caption{(Color online) Temporal evolution of the vortex number. (a) Decay curves obtained for various condensate fractions $\eta$. Solid lines are fitting curves from the rate equation in Eq~(1). The inset shows the atom number of the condensate, $N_0$, for each $\eta$. The data points for $\eta=0.80$ are identical to those in Fig.~4(a).  Decay rates (b) $\Gamma_1$ and (c) $\Gamma_2$ as a function of the sample temperature $T$. Solid lines are power-law fits to the data with exponents
of 0.87(20) and 1.88(12) for $\Gamma_1$ and $\Gamma_2$, respectively.}
\label{decay curve}
\end{figure}

We first investigate the temperature dependence of the decay rates. In order to reduce the effects from the variation of $\mu$, which determines the radial extent of the condensate, $R$, as well as the vortex core size $\xi$, the atom number of the condensate, $N_0=\eta N$, is kept to be almost constant [Fig.~\ref{decay curve}(a) inset], where $\eta$ is the condensate fraction and $N$ is the total atom number. In the Thomas-Fermi approximation, $\mu=(15 N_0 a/ \bar{a})^{2/5} \hbar \bar{\omega}/2=\frac{1}{2}m\omega_r^2 R^2$ with $a$ being the scattering length of atoms, $\bar{a}=\sqrt{\hbar/m\bar{\omega}}$, and $\bar{\omega}=\omega_r^{2/3}\omega_z^{1/3}$. $N$ and $\eta$ are measured right after turning off the repulsive laser beam. The temperature is estimated from the mean-field relations, $T=T_c(1-\eta)^{1/3}$ and $k_B T_c=0.94\hbar \bar{\omega} N^{1/3}$. During 15~s hold time, the condensate fraction $\eta$ was observed to decrease less than 10\%, showing that there is no significant heating in the relaxation. The lifetime of the sample in the trap was over 60~s. 

The measurement results are displayed in Fig.~\ref{decay curve}. Both of the decay rates monotonically increase as the temperature is increased [Fig.~\ref{decay curve}(b) and \ref{decay curve}(c)], demonstrating the thermal nature of the relaxation dynamics. It is noticeable that the two-body decay rate, $\Gamma_2$, shows a faster response to the temperature than the one-body decay rate, $\Gamma_1$, and this seems to imply that the physical mechanisms determining each decay rate are different. The power-law fits to $\Gamma_1(T)$ and $\Gamma_2(T)$ give the exponents of 0.87(20) and 1.88(12), respectively, showing almost linear and quadratic dependence on the temperature. Here, we assume that the decay rates vanish at $T=0$. The turbulent condensate may dynamically relax even at $T=0$, i.e. without thermal atoms, but in the temperature range of our measurements, $k_B T/\mu>2$, we presume that thermal dissipation effects primarily govern the decay dynamics.

The one-body decay rate $\Gamma_1$ is predominantly determined from the decay behavior at low $N_v$ which is mainly driven by the vortex drifting-out process. Thus, we might regard $\Gamma_1$ as the thermal damping rate of a trapped BEC containing vortices, in particular, with zero net vorticity. In previous theoretical studies on nonequilibrium dynamics of trapped BECs, linear $T$ dependence of the thermal damping rate was anticipated for low-energy excitations~\cite{FSW98} and vortex lattice formation~\cite{Gardiner,KTU}. However, its applicability to our measurements will require further theoretical work. 

\subsubsection{Chemical potential}

The vortex core size $\xi\propto\mu^{-1/2}$ not only provides a characteristic length scale in the vortex dynamics but also defines the lower bound for the separation of two distinguishable vortex cores. The two-body decay rate $\Gamma_2$ is associated with many-vortex effects including the vortex pair annihilation process and thus, it would be significantly affected by a change of the chemical potential. 

\begin{figure}
\includegraphics[width=8.5cm]{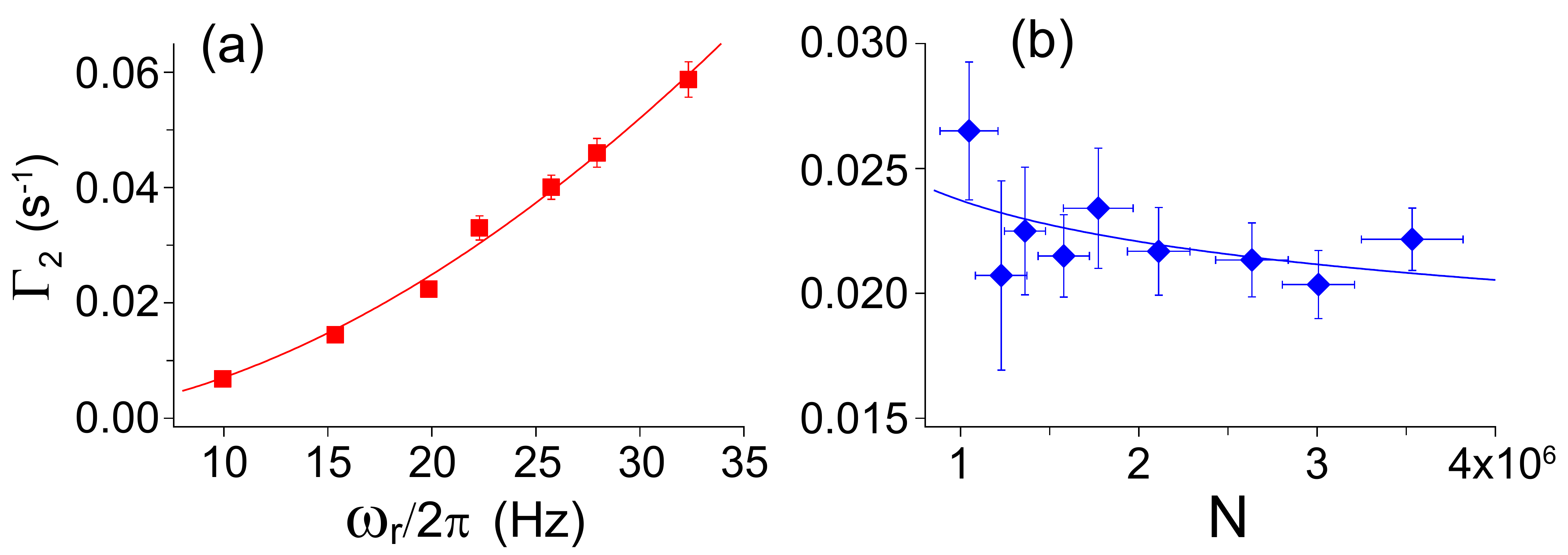}
\caption{(Color online) Decay rate $\Gamma_2$ as a function of (a) the radial trapping frequency $\omega_r$ ($\eta\approx 0.80$ and $N\approx 2.4\times10^6$) and (b) the total atom number $N$ ($\eta\approx0.63$ and $\omega_r/2\pi=15$~Hz). The power-law fits (solid lines) give the exponents of (a) 1.82(9) and (b) $-$0.10(6) for $\omega_r$ and $N$, respectively.}
\label{atomfreq}
\end{figure}

To investigate the $\mu$ dependence of $\Gamma_2$, we make two additional sets of measurements of $\Gamma_2$ for the variations of the trapping frequency $\omega_r$ and the total atom number $N$, respectively. In each measurement set, the condensate fraction $\eta$ is controlled to be fixed within a few \%. For varying $\omega_r$, we adiabatically ramp the magnetic trapping potential during the repulsive laser beam being turned off. From the mean-field relations, $T\propto (\omega_r^{2} N)^{1/3}$ and $\mu \propto (\omega_r^{2} N)^{2/5}$ for a fixed $\eta$. The change in $\mu$ entails a change in the spatial extent of the condensate as $\mu=\frac{1}{2}m\omega_r^2 R^2$. To decouple this effect, we define a local two-body decay rate as $\gamma_2 \equiv (\pi R^2) \Gamma_2$, crudely neglecting the details of the sample inhomogeneity. If $\gamma_2$ is proportional to $T^\alpha \mu^\beta$, then we would have $\Gamma_2 \propto \omega_r^{2+2\kappa/3} N^{\kappa/3}$, where $\kappa=\alpha+\frac{6}{5}(\beta-1)$.

\begin{figure}
\includegraphics[width=7.7cm]{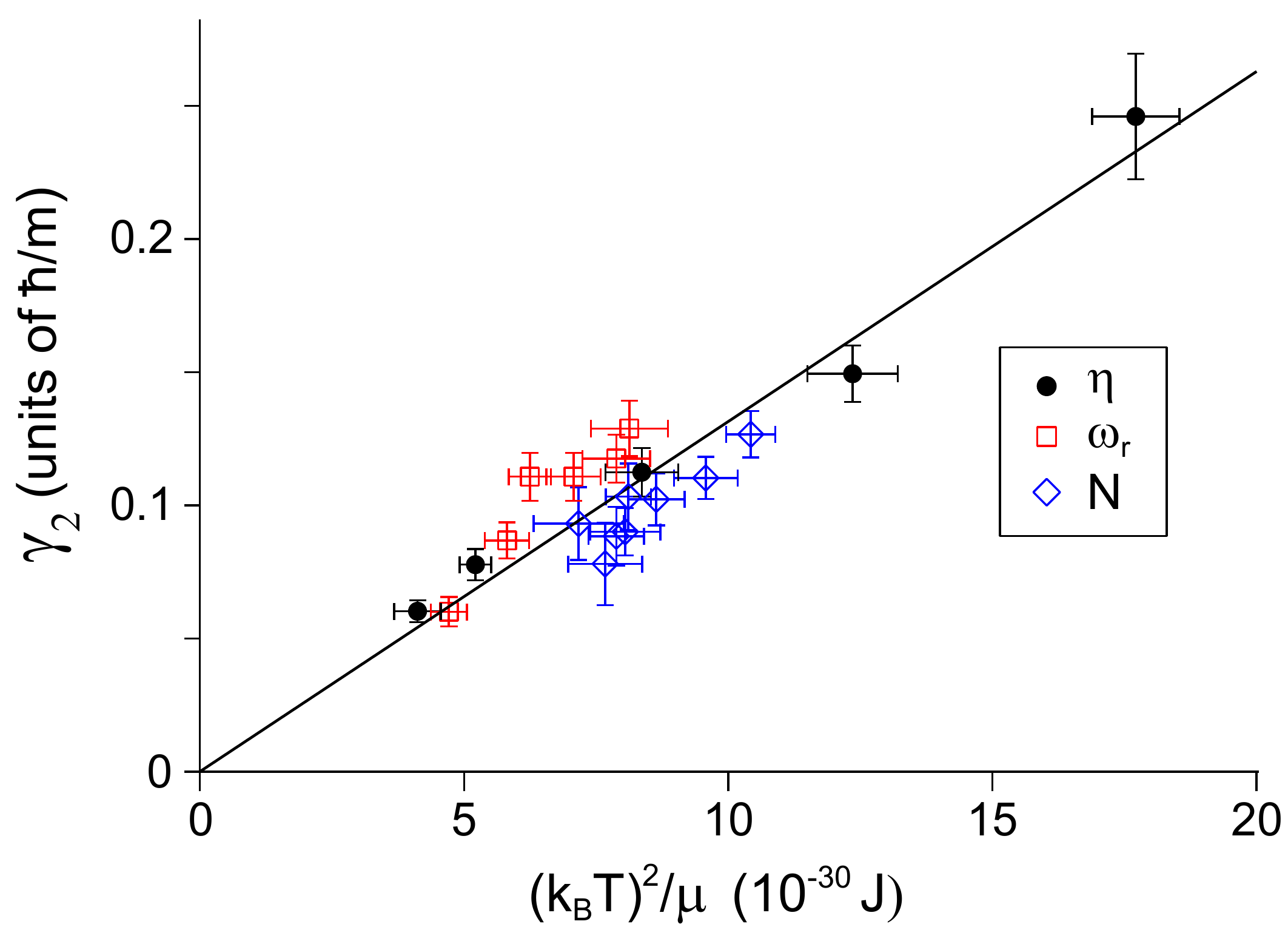}
\caption{(Color online) Local two-body decay rate $\gamma_2=(\pi R^2) \Gamma_2$ versus $(k_B T)^2/\mu$. The solid line denotes a linear fit to the data. Black solid circles, red open squares and blue open diamonds correspond to the $\Gamma_2$ data points in Fig.~7(c), Fig.~8(a), and Fig.~8(b), respectively.}
\label{localgamma}
\end{figure}

Figure~\ref{atomfreq} shows the decay rate $\Gamma_2$ as functions of $\omega_r$ and $N$. The power-law fits to the data gives $\Gamma_2 \propto \omega_r^{1.82(9)} N^{-0.10(6)}$ and the obtained exponents for $\omega_r$ and $N$ correspond to $\kappa=-0.27(14)$ and $-0.30(18)$, respectively. It is remarkable to observe that the two measurement sets give a consistent result. With the previous result $\alpha\approx 1.9$, $\kappa\approx -0.3$ suggests $\beta\approx -0.8$. In Fig~\ref{localgamma}, we plot all the $\Gamma_2$ measurements in the plane of $\gamma_2$ and $(k_B T)^2/\mu$ and see that they collapse fairly well in a line.

Our analysis shows that the local two-body decay rate is a useful quantity for characterizing the relaxation of our system. An important but still open question is what is the exact physical mechanism determining $\gamma_2$ in the relaxation dynamics. Let us consider the case where the two-body decay term in the rate equation purely originates from the vortex pair annihilation, assuming the vortex collision rate $\gamma_c=(\hbar/2m)n_v$ as discussed for the system at high $n_v$. From the relation $\Gamma_2 N_v= 2 p_a \gamma_c$, where $p_a$ is the annihilation probability for a vortex-antivortex collision event, the decay rate $\gamma_2$ would be expressed as $\gamma_2=(\hbar/m) p_a$. Here, the observation of $\gamma_2\propto T^2/\mu$ has an interesting implication that the dimensionless quantity $p_a$ is not determined as a function of the reduced temperature $\tilde{T}=k_B T/\mu$. This means that the vortex pair annihilation dynamics in our system cannot be explained with a pure 2D model which intrinsically preserves $\tilde{T}$-scaling behavior. The vortex line excitations, although their thermal excitations are suppressed, might play a role in the annihilation dynamics. Or the inhomogeneity of the trapped sample might be involved in a more intricate manner.

\section{Summary}

In summary, we have investigated thermal relaxation of superfluid turbulence in highly oblate Bose-Einstein condensates and presented possible evidences on the vortex-antivortex annihilation. We have characterized the relaxation of the turbulent condensate with the one-body and two-body decay rates of the vortex number. Our measurement results on the decay rates should provide a quantitative test on finite-temperature theories for vortex dynamics~\cite{HV,thermalQT,vring}. One interesting extension of this work would be exploring the crossover regime from 3D to 2D by increasing the axial confinement~\cite{nowak,phasefluc}. Even in a quasi-2D superfluid with $\mu<\hbar\omega_z$, thermal phase fluctuations might qualitatively modify the relaxation behavior of quantum turbulence.

\begin{acknowledgements}

We thank A.~J.~Allen and H.~Zhang for useful discussions. This work was supported by the NRF of Korea (Grant Nos.~2011-0017527, 2008-0062257, and 2013-H1A8A1003984).\\

\end{acknowledgements}

\textit{Note added in revision}: A numerical study on our experiment has been recently reported~\cite{stagg} and it reveals that the nonexponential decay of the vortex number originates from the pair annihilation in the turbulent condensate.

\end{document}